\documentclass[iop]{emulateapj}
\usepackage{natbib}
\usepackage[bookmarks=False]{hyperref} 
\usepackage{amsmath,amstext}
\usepackage[export]{adjustbox}
\usepackage[export]{adjustbox}
\usepackage[all]{hypcap} 

\usepackage{apjfonts} 
\newcommand{\Spitzer}{{\sl Spitzer }}

\shorttitle{Spitzer Decorrelation Methods}
\shortauthors{Kilpatrick et al.}

\begin{document}

   \title{Spitzer Secondary Eclipse Depths with Multiple Intrapixel Sensitivity Correction Methods\\ Observations of WASP-13b, WASP-15b, WASP-16b, WASP-62b, and HAT-P-22b}

   \author{Brian M. Kilpatrick\altaffilmark{1}, 
   Nikole K. Lewis\altaffilmark{2}, 
   Tiffany Kataria\altaffilmark{3}, 
   Drake Deming\altaffilmark{4}  ,
   James G. Ingalls \altaffilmark{5},
   Jessica E. Krick\altaffilmark{5}, 
   Gregory S. Tucker\altaffilmark{1} }

\altaffiltext{1}{Department of Physics, Box 1843, Brown University, Providence, RI 02904, USA; brian\_kilpatrick@brown.edu}
\altaffiltext{2}{Space Telescope Science Institute, Baltimore, MD 21218, USA; nlewis@stsci.org}
\altaffiltext{3}{Jet Propulsion Laboratory, California Institute of Technology, 4800 Oak Grove Drive, Pasadena, CA 91109, USA; tiffany.kataria@jpl.nasa.gov}
\altaffiltext{4}{Department of Astronomy, University of Maryland, College Park, MD 20742, USA; ddeming@astro.umd.edu}
\altaffiltext{5}{Spitzer Science Center, Infrared Processing and Analysis Center, California Institute of Technology, Mail Code 220-6,
Pasadena, CA 91125, USA; krick@ipac.caltech.edu}

\vspace{0.5 in}

\begin{abstract}

We measure the 4.5 $\mu$m thermal emission of five transiting hot Jupiters, WASP-13b, WASP-15b, WASP-16b, WASP-62b and HAT-P-22b using channel 2 of the Infrared Array Camera (IRAC) on the {\sl Spitzer Space Telescope}.  Significant intrapixel sensitivity variations in \Spitzer IRAC data require careful correction in order to achieve precision on the order of several hundred parts per million (ppm) for the measurement of exoplanet secondary eclipses.  We determine eclipse depths by first correcting the raw data using three independent data reduction methods.  The Pixel Gain Map (PMAP), Nearest Neighbors (NNBR), and Pixel Level Decorrelation (PLD) each correct for the intrapixel sensitivity effect in \Spitzer photometric time-series observations.   The results from each methodology are compared against each other to establish if they reach a statistically equivalent result in every case and to evaluate their ability to minimize uncertainty in the measurement.  We find that all three methods produce reliable results.  For every planet examined here NNBR and PLD produce results that are in statistical agreement.  However, the PMAP method appears to produce results in slight disagreement in cases where the stellar centroid is not kept consistently on the most well characterized area of the detector.  We evaluate the ability of each method to reduce the scatter in the residuals as well as in the correlated noise in the corrected data.  The NNBR and PLD methods consistently minimize both white and red noise levels and should be considered reliable and consistent.  The planets in this study span equilibrium temperatures from 1100 to 2000~K and have brightness temperatures that require either high albedo or efficient recirculation.  However, it is possible that other processes such as clouds or disequilibrium chemistry may also be responsible for producing these brightness temperatures.    \\

\end{abstract}
   \keywords{planets and satellites: atmospheres --
                planets and satellites: detection, techniques: photometric, methods:  numerical, atmospheric effects}

 \maketitle
%
\capstartfalse
\def\arraystretch{1.5}
\tabletypesize{\small}
\begin{deluxetable*}{lccccc}[!t]
\tablecaption{Observations\label{obstab}}
\tablehead{
\colhead{} & \colhead{WASP-13b} & \colhead{WASP-15b} &  \colhead{WASP-16b} & \colhead{WASP-62b} &\colhead{HAT-P-22b}} 
\startdata
AOR &45675520 & 45675776 & 45674496&48680448& 45674752\\
Date of Obs.(UT) &June 08, 2012 & September 14, 2012 & 	September 09, 2012 & September
19, 2013 &  June 03, 2012\\
Frame Time (s) & 0.4& 2.0 &  2.0 & 2.0 & 0.4\\
Duration of Obs. (min)& 469& 495 & 277 &  	444 &   371\\
\enddata
\end{deluxetable*}
\capstarttrue

\section{Introduction}\label{sec: Intro}
The {\sl Spitzer Space Telescope} has been the preeminent observatory used to obtain photometric light curves of transiting exoplanets in the infrared. The relative flux variations during a secondary eclipse (planet passing behind the star) provide insight into the planetary energy budget and atmospheric circulation.  Hot Jupiters, Jupiter-sized planets less than 0.1~AU from their host stars,  are valuable targets for such studies in the near infrared.  The \Spitzer Infrared Array Camera (IRAC) instrument has two channels well positioned to sample the peak of hot-Jupiter emission spectra. The decrease in relative flux during a secondary eclipse is representative of the planetary dayside emission.  The magnitude of this signal is often on the order of several hundred to a few thousand parts per million (ppm). In its post cryogenic `Warm Mission', {\sl Spitzer} IRAC is capable of obtaining better than 100 ppm precision in time series observations \citep{Ingalls}.  However, these IRAC observations are also affected by systematic and spatially correlated variations as the image centroid moves across a detector pixel.  The  intra-pixel sensitivity effect in the under-sampled camera can cause variations on the order of 10\% with normal pointing jitter and movement \citep{2016Ingalls}.  This movement is attributed to several factors but on timescales of the order of several hours is primarily due to jitter (high frequency and stochastic) and a heater cycling (wobble) that changes the alignment between the star tracker and optical axes by $\sim 0.15${\tt "} over a 40 minute period \citep{Grillmair}.  

There have been many methods used to remove this correlated noise effect \citep[e.g.][] {2005Reach, 2008Charb, 2010Ballard}, but the past several years have seen the development of a few novel methods that have been utilized in a number of recent \Spitzer publications \citep[e.g.][]{Lewis, Deming, stevenson2012, Ingalls, Gibson2012, Evans2015, Morello2015}.  

In this work we consider three of the more commonly used methods; the Infrared Processing and Analysis Center's (IPAC) provided Pixel Variation Gain Map (PMAP) \citep{Ingalls}, the Nearest Neighbor Method (NNBR) \citep{Lewis}, and Pixel Level Decorrelation (PLD) \citep{Deming}.  Each of these methods has been independently tested and used to analyze various data sets.  A recent IPAC Data Challenge \citep{2016Ingalls} invited members of the community to employ these methods, and others, in the reduction of a set of multi-epoch eclipses of the planet XO-3b along with a set of synthetic data.  The data challenge sought to show that the full range of reduction methodologies could provide accurate and consistent results over many observations of the same planet.  

The work reported here continues the effort to validate that each method produces results that are in statistical agreement with each other by using measurements of five different planets.  This approach will provide insight into how well each method performs in reducing correlated noise over a range of eclipse depths, observation times, cadence, and pointing stability.  Note for comparison purposes that in the data challenge and in \citet{2016Krick} NNBR is referred to as Kernel Regression with Data and what we will later define as KMAP is referred to as Kernel Regression with Pixel Map.

\section{Observations}

The observations analyzed here are all part of Program ID 80016 (PI: J. Krick) and include the planets WASP-13b, WASP-15b, WASP-16b, WASP-62b and HAT-P-22b. Each planet was observed during one secondary eclipse by IRAC Channel 2 ($4.5\, \mu{\rm m}$ bandpass) \citep{Fazio2004}.  The details of each Astronomical Observing Request (AOR) are displayed in Table \ref{obstab}.  All of these observations were carried out in sub-array mode ($32\times 32$ pixels, $39{\tt "} \times 39{\tt "}$) with a 30 minute peak-up observation preceding them.  The use of a peak up observation allows the instrument to stabilize the image on the detector `sweet spot' and decreases the likelihood of a ramp in the data \citep{Ingalls}.
  
\section{Methods}\label{Methods}

In each case we began with Basic Calibrated Data (BCD) available on the Spitzer Heritage Archive.  Each BCD file contains a cube of 64 frames of $64\times 64$ pixels.  Each frame was corrected by two separate methods for bad pixels (pixels with values outside of a pre-defined range, which we take to be ($-100$, $10,000$)) or NaN values.  The PMAP routine defines the area of an annulus with a 3 pixel inner radius and 7 pixel outer radius centered on the stellar centroid as the background.

The PMAP photometry routine ignores bad pixels in the background when calculating the sky contribution but will not produce a flux value from aperture photometry if a bad pixel is found within the 3 pixel aperture radius.

NNBR and PLD employ a slightly different photometry routine which replaces any bad pixels or NaN values outside of the aperture with the median background value.
The background is defined as any point outside a 10 pixel radius from the stellar centroid.   All points in the background area were sorted, clipped at $3\sigma$ three times to remove outliers, then fit with a Gaussian to determine the sky value and uncertainty in the background subtraction.

Time-series photometry data were filtered to remove outliers by iteratively clipping values outside $3\sigma$ of the median of the nearest 50 points temporally. Less than 1\% of the data was removed by this filtering. Each eclipse fit was based on the model of \cite{Mandel} for a uniform occultation.  
\Spitzer IRAC data is known to have an exponential ramp in flux over the first 30--60  minutes of observing \citep[e.g.][]{Lewis, Knutson2012}; however the peak-up technique has alleviated this problem to some extent.  As a precaution, each data set was trimmed at 10, 20, and 30 minutes from the beginning to see if this resulted in a decrease in the standard deviation of the normalized residuals (SDNR).  In each case we determined that trimming was not necessary.  Aside from these commonalities, the particulars of each method are described in the following sections.

\subsection{Pixel Gain Map}\label{ssec: pmapsec}

The PMAP method was applied using the tools available from the IRAC program website \footnote{\label{iracweb} \url{http://irachpp.spitzer.caltech.edu/page/contrib}}.  The instructions provided were followed closely in an attempt to produce consistent results.  BCD files from the IRAC Data Reduction Pipeline were downloaded and analyzed as follows.  The IDL {\tt box\_centroider.pro} routine 
 calculates the centroid (x,y) position as shown in \citet{Ingalls}:  
\begin{equation}
x_{\rm cen}=\frac{\sum_{j,k}(I_{jk}j)}{\sum_{j,k}I_{jk}}; \qquad
y_{\rm cen}=\frac{\sum_{j,k}(I_{jk}k)}{\sum_{j,k}I_{jk}}.
\end{equation}
Here, $I_{jk}$ is the surface brightness of pixel (j,k), where the center of bottom left pixel of the subarray is position (0,0) \citep{Ingalls}.  In practice, we confine the centroiding to a 7 $\times$ 7 pixel box  with the pixel containing the peak flux at the center. 
Based upon the recommendations from IPAC
\footnote{\label{ipac} \url{http://irsa.ipac.caltech.edu/data/SPITZER/docs/dataanalysistools/tools/contributed/irac/iracpc_pmap_corr/}}
, we used a fixed radius circular aperture of 3 pixels.  The IDL routine {\tt aper.pro} was used to integrate over a circular area of square pixels.  The filtered flux along with the corresponding $x$ and $y$ positions of the stellar centroid were passed to the {\tt iracpc\_pmap\_corr.pro} IDL routine in order to calculate the corrected flux values.  The derivation of the photometric gain maps is discussed in detail in \cite{Ingalls}. The peak of pixel sensitivity for channel 2 corresponds to a position of (15.12, 15.09).  The sweet spot is a $0.5 \times 0.5$ pixel box centered at the position of peak sensitivity.

\begin{figure}
\centering
\includegraphics[trim=2.0in 3.4in 2.0in 3.4in,clip,width=0.48\textwidth]{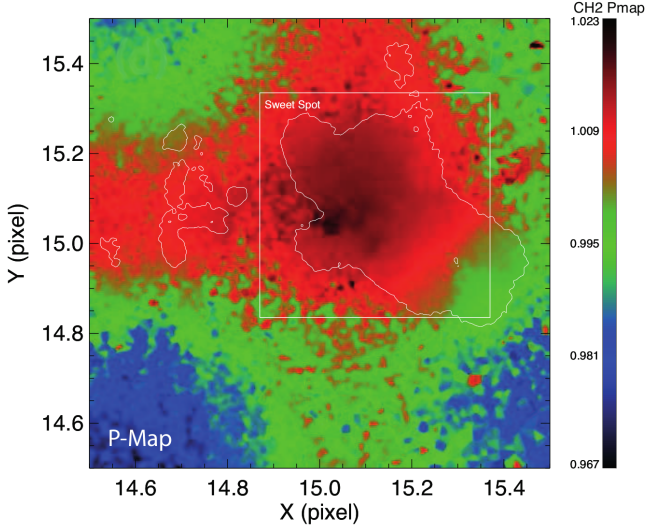} 
\caption{This plot from \cite{Ingalls} shows how the pixel responds as a function of stellar centroid position.  The sweet spot is a $0.5 \times 0.5$ pixel area outlined by the white box in the figure.  The color map shows the pixel response, or gain, at each location on the pixel.  The areas enclosed by the contour lines are areas with an occupation number of 20 or greater.   For the most reliable results the observation should seek to keep the stellar centroid in an area that is both highly responsive and well characterized (i.e. has a high occupation number).}\label{pmapfig}
\end{figure}

A non-variable calibration target was observed at various offsets from the peak in order to create the grid of relative flux values.  The ch2 gain map has 409, 539 photometry points in it, 90\% of which are within the sweet spot.  Each point on the grid is the result of a combination of a number of observational points.  This number, the occupation number, can be used to assess the reliability of any point on the grid.  Figure \ref{pmapfig}
shows occupation number contours.  Areas outside of the contours do not have a high enough occupation number to be considered accurate. 

Once the flux values were corrected, a Levenberg-Marquardt (LM) fitting routine \citep{Markwardt2009} determined the best fit for eclipse depth, center of eclipse, nd stellar flux baseline using the  \cite{Mandel} model for a uniform occultation.

Values for $\frac{a}{R_\star}$, $\frac{R_p}{R_\star}$, and inclination (listed in Table \ref{ephem}), from the exoplanets.org database \citep{exodat} were input as constants in the calculation of the eclipse model.  The results of the LM minimization were used to seed a multi-chain Markov-Chain Monte Carlo (MCMC) simulation to determine the best fit and uncertainties for the eclipse depth and time of eclipse parameters.  \\

In order to take correlated ``red" noise into account in our uncertainties we binned the residuals from the LM fit in bin sizes of 2.0 s intervals up to 90 min and calculated the $\beta_{\rm red}$ coefficient as defined in \cite{Gillon2010}.  The maximum $\beta_{\rm red}$ value over the entire 90 min range was used as the worst case scenario.  The uncertainty of each photometric point, as determined by the SDNR of the unbinned time-series, was multiplied by the $\beta_{\rm red}$ factor before being passed to the MCMC to account for time-correlated noise. The priors for each parameter in the MCMC were based on a normal distribution centered at the LM result with a width determined by the uncertainty in the LM result.

We employ a Metropolis-Hasting algorithm within the Gibbs sampler MCMC \citep{Ford2005}.  There were two free parameters, depth and center of eclipse.  Four independent chains were initiated from four unique and randomly selected starting points, run for a minimum of $10^4$ steps and then tested for Gelman-Rubin convergence \citep{gelman1992}.  The algorithm sought to minimize $\chi^2$.  Step size was adjusted to produce $\sim$ 40\% acceptance rate.  The first 20\% of the steps were discarded to remove any burn in period and the rest were kept to create a histogram of results for each parameter. After inspecting each histogram for evidence of asymmetry; we determined that symmetric uncertainty would be appropriate.  We find the best fit Gaussian to the MCMC histogram and take the $1\sigma$ width of the Gaussian to be the $1\sigma$ uncertainty in the parameter.\\

\subsection{Nearest Neighbors}\label{ssec: nnbrsec}

Each data set was also corrected for intra-pixel sensitivity variations using the NNBR as applied to HAT-P-2b phase-curve observations in \cite{Lewis}.  Using this method we take the BCD files and calculate the centroid position of each exposure using a center of light method.  The noise pixel parameter ($\tilde{\beta}$), defined by equation A2 of \cite{Lewis} as
\begin{equation}\tilde{\beta}=\frac{\left(\sum P_i\right)^2}{\sum \left(P_i^2\right)}, \end{equation} 
is calculated for each exposure. Circular aperture photometry is performed using fixed and variable aperture radii.  
 Variable radius apertures were based on either multiplying $\tilde{\beta}$ by some scaling factor or adding/subtracting some constant from $\tilde{\beta}$.  
Each flux value $i$ was linked with its 50 nearest neighbors $j$ by distance:
\begin{equation}
 r_{i,j}=\sqrt{a(x_i-x_j)^2+b(y_j-y_i)^2+c(\tilde{\beta}_j^{\frac{1}{2}}-\tilde{\beta}_i^{\frac{1}{2}})^2}.
 \end{equation}
 
Each nearest neighbor was weighted with a Gaussian smoothing kernel $K_i(j)=$
  \begin{equation}
  exp\left(-\frac{(x_j-x_i)^2}{2\sigma^2_{x,i}}-\frac{(y_j-y_i)^2}{2\sigma^2_{y,i}}-\frac{(\tilde{\beta}_j^{\frac{1}{2}}-\tilde{\beta}_i^{\frac{1}{2}})^2}{2\sigma^2_{\tilde{\beta}_j^{\frac{1}{2}},i}}\right),
\end{equation}
where $x$ and $y$ are the stellar centroid locations on the detector and the three-dimensional widths of the smoothing kernel are controlled by the $\sigma$ terms that adjust depending on the density of the nearest neighbors (see \citet{Lewis} for further details).

The corrected relative flux value for any photometric point $F_i$ with centroid position ($x_i$, $y_i$) and uncorrected flux value $F_{0,i}$ then becomes
\begin{equation}
F_i=\frac{F_{0,i}}{W_i(x_i, y_i)},
\end{equation}
where
\begin{equation}
W_i(x_i, y_i)=\frac{\sum_jF_{0,j} K_{i}(j)}{\sum_jK_{i}(j)}.
\end{equation}
$W_i$(x$_i$, y$_i$) is summed over the 50 nearest neighbors (those with the smallest $r_{i,j}$ values).  Best fit parameters for each aperture were solved for using the same LM minimization routine as described in \autoref{ssec:  pmapsec}.  The fit from each aperture size, including both fixed and variable, was evaluated for goodness of fit by considering the SDNR and the  maximum $\beta_{\rm red}$ factor over the same range as PMAP.  The aperture which minimized both $\beta_{\rm red}$ and SDNR was chosen and sent to the four chain MCMC to fine tune parameter fits and determine uncertainties.  Error on the data points used in the MCMC are equivalent to the SDNR multiplied by the $\beta_{\rm red}$  factor.  The aperture type and radius (pixels) chosen for each planet are as follows:  Var. 1.98, Fixed 2.45, Var. 2.70, Var. 2.88, Fixed 2.00 for WASP-13b, WASP-15b, WASP-16b, WASP-62b, and HAT-P-22b respectively.  In cases where a variable aperture was used, the radius presented is the mean radius over the entire data set.   MCMC analysis as described in \autoref{ssec: pmapsec} was used to establish uncertainty in the parameter results.  In this case there are still only the same two free parameters, however, it is important to note that the correction applied to the raw data (the nearest neighbor map) is dependent upon these two parameters and recalculated at each step in the MCMC.  Conversely the PMAP is independent of choices of eclipse depth and center.

\capstartfalse
\def\arraystretch{1.5}
\tabletypesize{\footnotesize}
\begin{deluxetable*}{lccccc}[!t]
\tablecaption{Ephemerides\label{ephem}}
\tablewidth{0.95\textwidth}
\tablehead{
\colhead{Parameter} & \colhead{WASP-13b} & \colhead{WASP-15b} &  \colhead{WASP-16b} & \colhead{WASP-62b} &\colhead{HAT-P-22b}} \label{ephem}
\startdata
$T_{\star}$ &5950$\pm$70  & 6300$\pm$100 & 5700$\pm$150&6230$\pm$80 & 5302$\pm$80\\
$M_\star$ (M$_{\rm Sun})$ &1.090$\pm$0.05  & 1.180$\pm$0.03 & 1.000$\pm$0.0.03&1.250$\pm$0.05 & 0.916$\pm$0.35\\
$R_\star$ (R$_{\rm Sun})$ &1.574$\pm$0.048  & 1.477$\pm$0.072 & 0.946$\pm$0.0.057&1.280$\pm$0.05 & 1.040$\pm$0.044\\
$M_p$ (M$_{\rm Jup})$ &0.474$\pm$0.034  & 0.543$\pm$0.021 & 0.842$\pm$0.0.032&0.562$\pm$0.042 & 2.151$\pm$0.077\\
$R_p$ (R$_{\rm Jup})$ &1.407$\pm$0.052  & 1.379$\pm$0.067 & 1.008$\pm$0.0.083&1.390$\pm$0.060 & 1.080$\pm$0.058\\
$R_p/R_{\star}$ &0.0919$\pm$0.0126  & 0.0984$\pm$0.0114 & 0.1095$\pm$0.0228&0.1109$\pm$0.014 & 0.1063$\pm$0.06\\
log(g)&2.775$\pm$0.042  & 2.829$\pm$0.046 & 3.315$\pm$0.055&2.865$\pm$0.047 & 3.660$\pm$0.144\\
Period & 4.3530135$\pm$2.7$\times10^{-6} $& 3.752100$^{+9\times10^{-6}}_{-1.1\times10^{-5}}$   &  3.118601$\pm$1.46$\times10^{-5}$ & 4.4119530 $\pm3\times10^{-6}$ & 3.2122200$\pm$9$\times10^{-6}$\\
$i$ ($^{\circ}$) & 85.43$\pm$0.29 & 85.96$^{+0.29}_{-0.41}$ & 85.22$\pm$0.35 &  	88.30 $^{+0.9}_{-0.6}$ &   86.90 $^{+0.6}_{-0.5}$\\
m$\sin{i}$ &0.472$\pm$0.034  & 0.541$\pm$0.021 & 0.839$\pm$0.032&0.562$\pm$0.042 & 2.148$\pm$0.077\\
$a/R_{\star}$  &7.35$\pm$0.26 & 7.29$\pm$0.38  & 	9.52$\pm$0.57 & 9.55$\pm$0.41 &  8.58$\pm$0.39\\
$T_c$ \tablenotemark{a} & 55575.5136$\pm$0.00160 & 54584.69819$^{+0.00021}_{-0.0002}$ &54584.42878$^{ +0.00035}_{-0.00025}$ &  55855.39195 $\pm$0.00027 & 54930.22001$\pm$0.00025\\
$T_{14}$ (d)\tablenotemark{a} & 0.1693$^{+0.00108}_{-0.00133}$ &  0.1813$\pm$0.0013 & 0.0800$\pm$0.0018 & 0.1588$\pm$0.0014 & 0.1196$\pm$0.0014\\   
\enddata
\tablenotetext{a}{$T_{14}$ is the total transit or eclipse duration. }
\end{deluxetable*}
\capstarttrue

\subsection{Pixel Level Decorrelation}

Using the PLD method of \cite{Deming}, we apply an array of both fixed and variable circular apertures (as a function of $\tilde{\beta}$) to calculate photometric flux values from BCD data.  The array of $3 \times 3$ pixels centered on the stellar centroid were saved and normalized so that at any time step their sum was unity thus removing any astrophysical signal.   PLD assumes that the total flux observed can be broken down into a linear equation:

\begin{equation} \label{seven}
\Delta S^{t} = \sum \limits _{i=1}^n c_i\hat{P}^t_i +DE(t) +ft+gt^2+h,
\end{equation}
where the pixel value of each of the $3 \times 3$ saved pixels is multiplied by some coefficient ($c_i\hat{P}^t_i$) and summed along with the eclipse model ($DE(t)$)and a time dependent `ramp' ($ft+gt^2$).
We introduce the eclipse through the \cite{Mandel} model.  This is the $E(t)$ term in the equation, and the free parameter $D$ is the eclipse depth.  This model component is then normalized by
\begin{equation}
{\rm model}=({\rm model}-1.0)/(1.0-{\rm min(model)}).
\end{equation}
The result is a light curve that is zero out of eclipse and $-1$ in eclipse. The problem is then reduced to solving the linear equation (Equation \ref{seven}) using matrix inversion.

A regression routine solves the equation for the coefficients over an array of several hundred different values for the center of eclipse.  The position of the center of eclipse which produces the smallest $\chi^2$ value is kept as the solution for the center of eclipse.  

One of the key components to the PLD method, in its attempt to reduce red noise, is to bin the data over various time scales and to find a solution at each binning.  The solution found at each binning is then applied to the full set of unbinned data and evaluated for goodness of fit by determining the SDNR and maximum $\beta_{\rm red}$.  The fit which minimizes these noise components is selected as the best fit for that aperture.  This process is repeated over each aperture size until a best fit is found for each.  

Once the best fit for each aperture is found we follow the broadband solution method outlined in \cite{Deming}.  Each set of residuals is binned at various timescales up to ingress/egress timescales.  The logarithm of the standard deviation of the residuals at each bin size is plotted against the logarithm of the bin size in time (s).  A line of slope $-0.5$ passing through the log of the unbinned SDNR is used as the theoretical solution.  The aperture that minimizes the $\chi^2$ with the theoretical model is chosen as the best fit and sent to a MCMC to fine tune the solution and establish the uncertainties.  The $\beta_{\rm red}$ coefficient is again used as a multiplicative factor on the SDNR for the uncertainty in each time series data point in the MCMC.  A similar MCMC algortithm as described in \autoref{ssec: pmapsec} is used here with 14 free parameters (nine pixel coefficients, depth and center of eclipse, two time coefficients, and offset).  

\section{Results}

The secondary eclipse depth and time of the center of eclipse derived for each observation are shown in Table \ref{res}.  These results are derived by fitting a Gaussian to the histogram of results from the MCMC chains, minus a 20\% burn in period.  Visual inspection of the histograms indicate no asymmetry, so we assume the uncertainty in each measurement to be symmetric and equivalent to the $1\sigma$ value of the Gaussian distribution. The brightness temperature for each planet was calculated using the methods of \cite{Seagerbook}.  The stellar properties used are shown in Table \ref{ephem}.  

\capstartfalse
\def\arraystretch{1.5}
\tabletypesize{\footnotesize}
\begin{deluxetable*}{lccccc}
\tablecaption{Results\label{res}}
\tablewidth{0.99\textwidth}
\tablehead{
\colhead{Parameter} & \colhead{WASP-13b} & \colhead{WASP-15b} &  \colhead{WASP-16b} & \colhead{WASP-62b} &\colhead{HAT-P-22b}} \label{ephem}
\startdata
PMAP Depth (ppm) &570$\pm$155&932$\pm$230&1010$\pm$149 &1025$\pm$172 &1466$\pm$120  \\
NNBR Depth (ppm)&977$\pm$260 & 954$\pm$221  &1060$\pm$222  &761$\pm$312  &1018$\pm$88 \\
NNBR$_{{\rm fixed}}$ Depth (ppm)&1027$\pm$221 & 981$\pm$142  &1131$\pm$118  &874$\pm$156  &1003$\pm$106 \\
PLD Depth (ppm) &1261$\pm$123&832$\pm$179  &1075$\pm$265 & 882$\pm$95 	&1120$\pm$91   \\
PLD$_{{\rm fixed}}$ Depth (ppm) &1232$\pm$123&841$\pm$217  &1099$\pm$233 & 948$\pm$103 	&1178$\pm$77   \\ 
 &  &   &  &  & \\
PMAP $T_c$  \tablenotemark{a} &1086.9964$\pm$0.0106 &1184.9675$\pm$0.0139  & 1179.5994$\pm$0.0056 & 1554.7008$\pm$0.0078 & 1081.8057$\pm$0.0029 \\
NNBR $T_c$ \tablenotemark{a} &1086.9977$\pm$0.0093  &1184.9644$\pm$0.0063 &1179.6017$\pm$0.0039& 1554.6958$\pm$0.0080 &1081.8041$\pm$0.0010  \\
NNBR$_{{\rm fixed}}$ $T_c$ \tablenotemark{a} &1086.9954$\pm$0.0110  &1184.9648$\pm$0.0060 &1179.6001$\pm$0.0032& 1554.6964$\pm$0.0060 &1081.8037$\pm$0.0033  \\
PLD $T_c$  \tablenotemark{a}&1086.9958$\pm$0.0022  &1184.9714$\pm$0.0066   &1179.6007$\pm$0.0055  &1554.6992$\pm$0.0029  &1081.8049$\pm$0.0013 \\   
PLD$_{{\rm fixed}}$ $T_c$ \tablenotemark{a}&1086.9963$\pm$0.0018  &1184.9704$\pm$0.0063   &1179.6001$\pm$0.0062  &1554.6992$\pm$0.0029  &1081.8044$\pm$0.0011   \\  
 &  &   &  &  & \\
PMAP $T_{B} ^\circ K$ &1253.0$\pm$123.7 &1485.04$\pm$151.0 & 1113.65 $\pm$54.3  & 1393.4$\pm$92  & 1518.9$\pm$52.1   \\
NNBR $T_{B} ^\circ K$ & 1581.7$\pm$151.0  & 1499.42$\pm$144  &1131.71$\pm$79.4   & 1245.4$\pm$184.5 & 1313.9$\pm$42.8 \\   
PLD $T_{B} ^\circ K$ & 1732.6$\pm$77.4 &  1418.30$\pm$121.6 & 1137.1$\pm$94.3 & 1315.1$\pm$53.4 & 1362.7$\pm$42.9  \\ 
 &  &   &  &  & \\   
Photon Noise Limit  &0.0098 & 0.0057  & 0.0055 &0.0041  & 0.0056\\
 &  &   &  &  & \\   
PMAP $\sigma_{w}$  &0.0144 & 0.0069  & 0.0067 &0.0048  & 0.0076\\   
NNBR $\sigma_{w}$ & 0.0122 &0.0063   &0.0063  & 0.0045 &0.0066 \\   
PLD $\sigma_{w}$ &0.0125  &0.0061   &0.0061  &0.0045  &0.0066 \\
 &  &   &  &  & \\   
PMAP $\beta_{{\rm red}}$  &1.269 & 2.457  & 1.183 &2.299 & 2.554\\  
NNBR $\beta_{{\rm red}}$  &2.112 & 1.673  & 1.081 &1.459  & 1.587\\   
PLD $\beta_{{\rm red}}$  & 1.071& 1.347  & 1.172 & 1.258 & 1.232\\    
 &  &   &  &  & \\   
PMAP $\sigma_{r}$  & 0.00049 & 0.00046  & 0.00036 & 0.00058 & 0.00074\\   
NNBR $\sigma_{r}$ &0.00061  &0.00046   &0.00037 & 0.00036 &0.00024 \\   
PLD $\sigma_{r}$ &0.00044  & 0.00039  &0.00020  &0.00036  &0.00023 \\ 
\enddata
\tablenotetext{a}{We list all center of eclipse times in BJD$\_$UTC-2.455E6 for consistency with other studies; to convert to BJD$\_$TT add 66.184 s.}
\end{deluxetable*}
\capstarttrue

\begin{figure*}[!h]\label{3plots}
\includegraphics[trim=0.0in 0.4in 0.0in 0.1in,clip,height=0.18\textheight,width=0.3\textwidth]{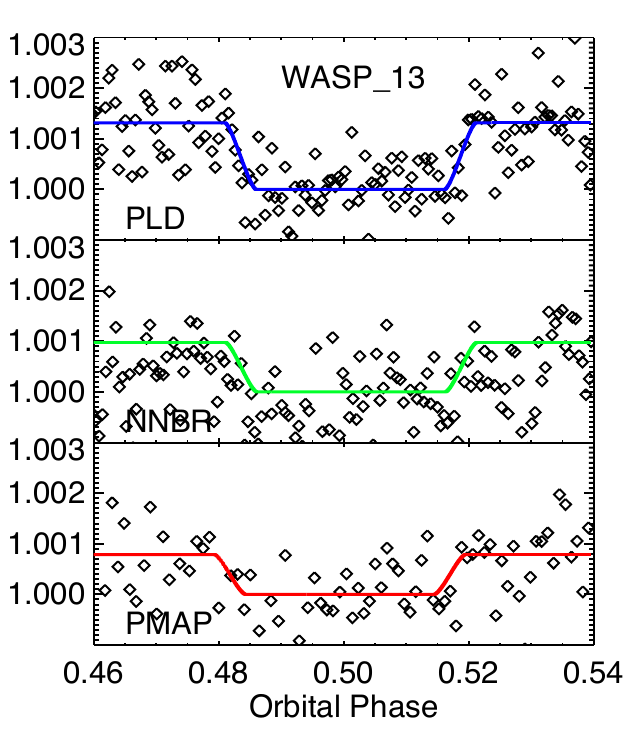} 
\includegraphics[trim=0.4in 0.33in 0.15in 0.35in,clip, height=0.18\textheight,width=0.3\textwidth]{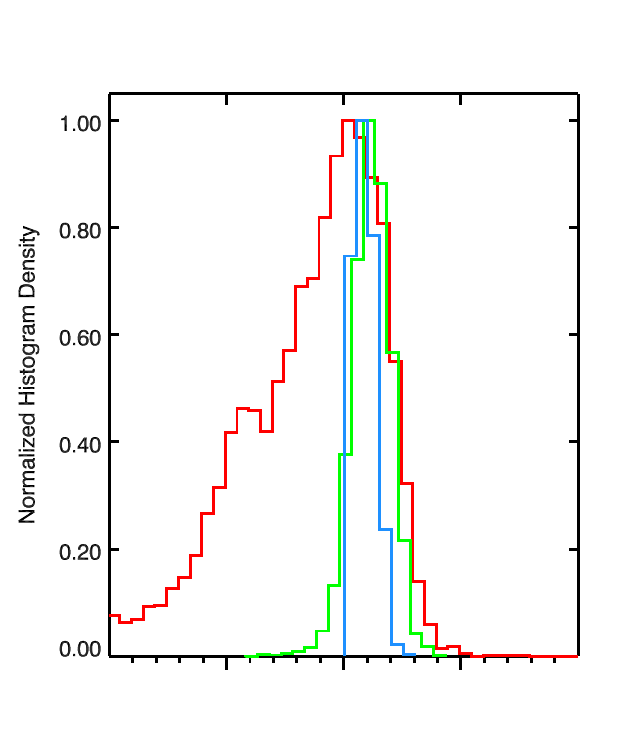}
\includegraphics[trim=0.4in 0.33in 0.0 0.35in, clip, height=0.18\textheight,width=0.3\textwidth]{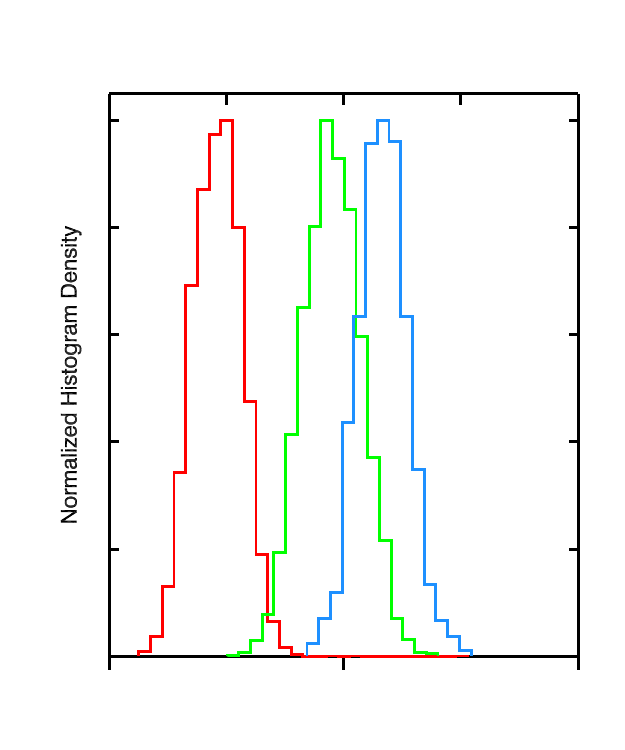}\\

\includegraphics[trim=0.0in 0.4in 0.0in 0.1in,clip,height=0.18\textheight,width=0.3\textwidth]{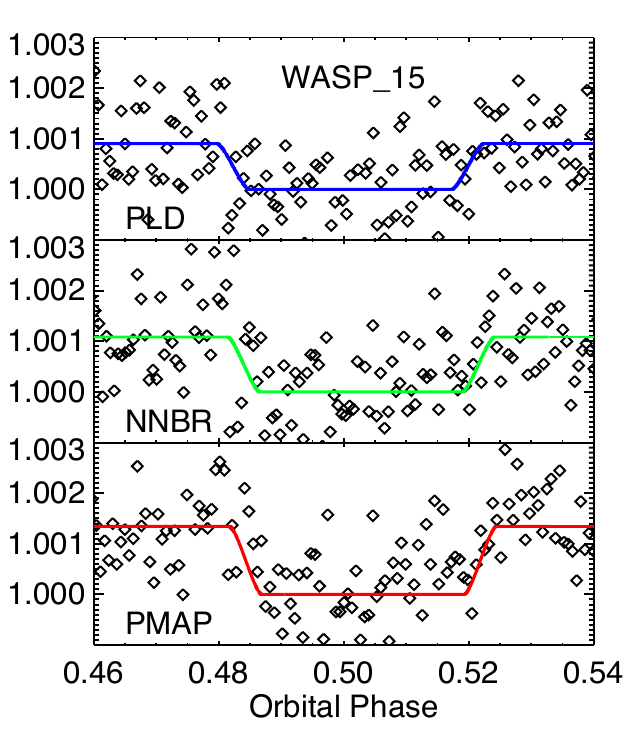} 
\includegraphics[trim=0.4in 0.33in 0.15in 0.35in,clip, height=0.18\textheight,width=0.3\textwidth]{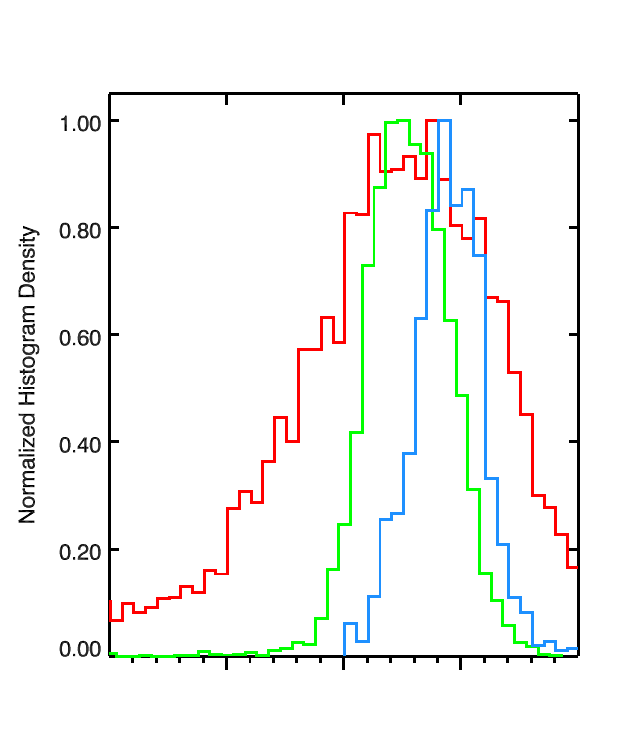}
\includegraphics[trim=0.4in 0.33in 0.0 0.35in, clip, height=0.18\textheight,width=0.3\textwidth]{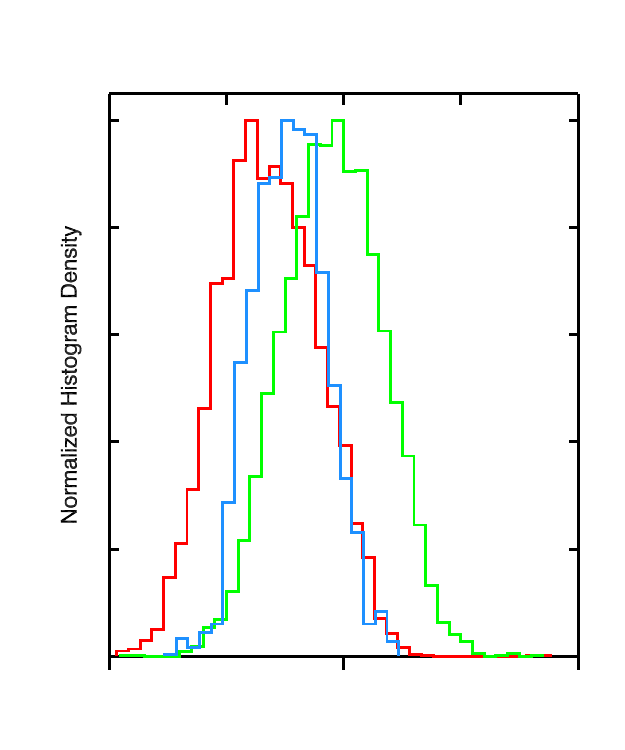}\\

\includegraphics[trim=0.0in 0.4in 0.0in 0.1in,clip,height=0.18\textheight,width=0.3\textwidth]{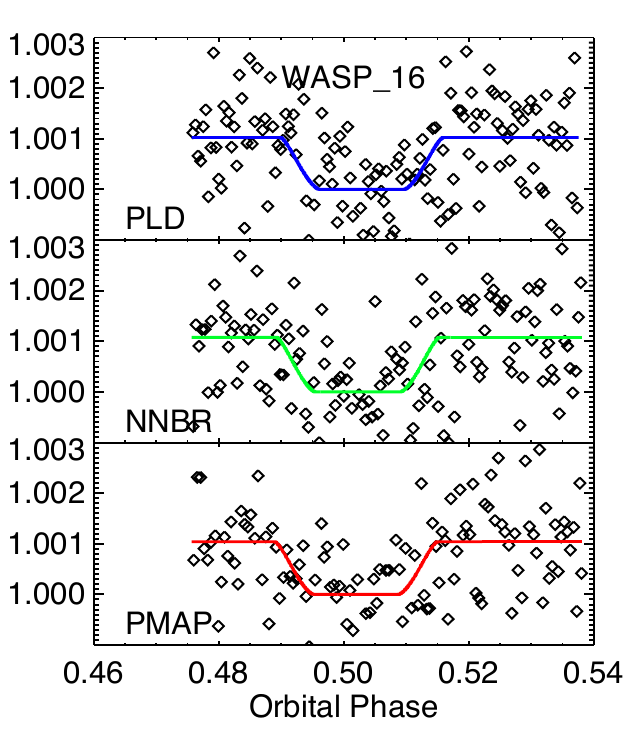} 
\includegraphics[trim=0.4in 0.33in 0.15in 0.35in,clip, height=0.18\textheight,width=0.3\textwidth]{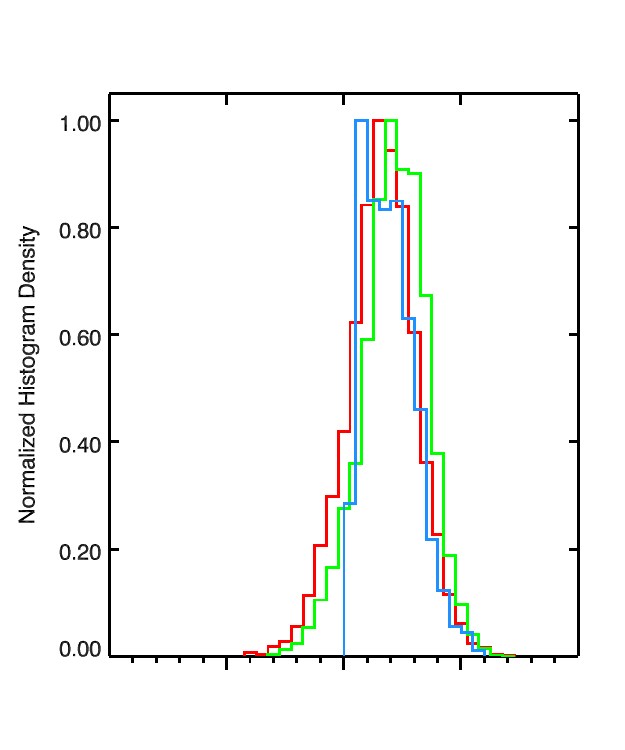}
\includegraphics[trim=0.4in 0.33in 0.0 0.35in, clip, height=0.18\textheight,width=0.3\textwidth]{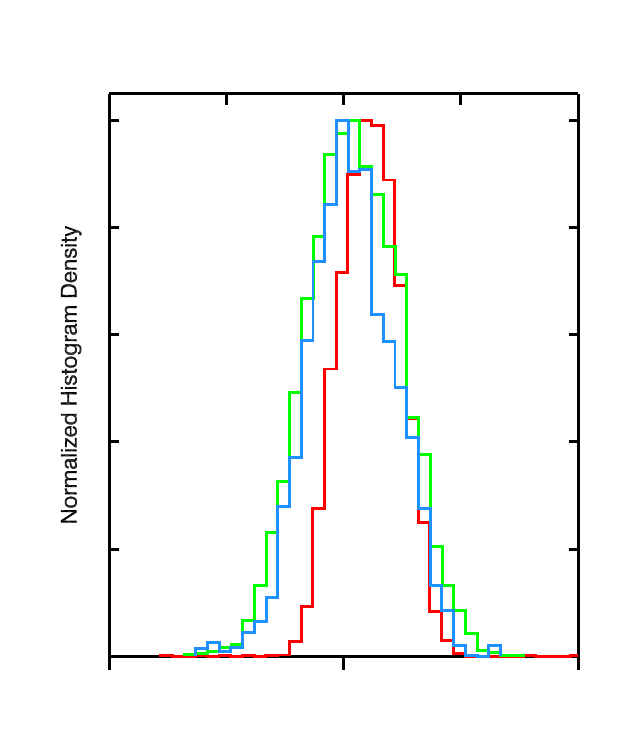}\\

\includegraphics[trim=0.0in 0.4in 0.0in 0.1in,clip,height=0.18\textheight,width=0.3\textwidth]{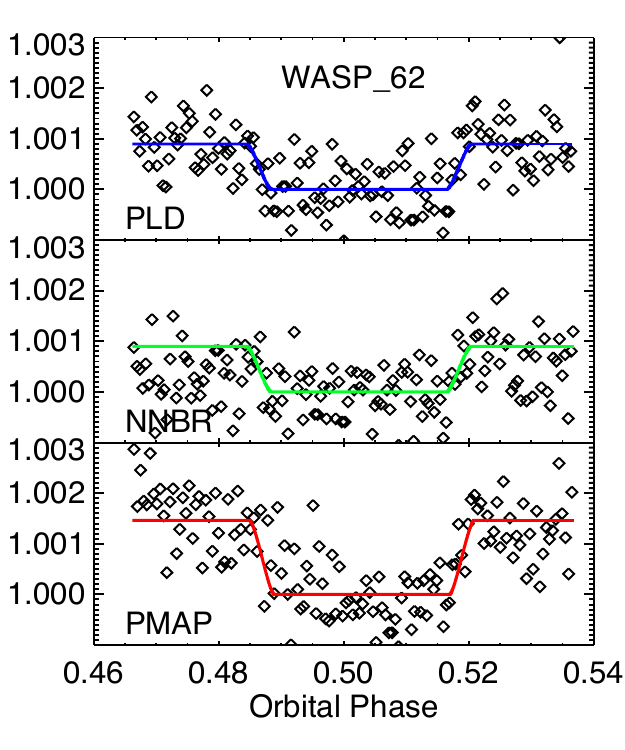} 
\includegraphics[trim=0.4in 0.33in 0.15in 0.35in,clip, height=0.18\textheight,width=0.3\textwidth]{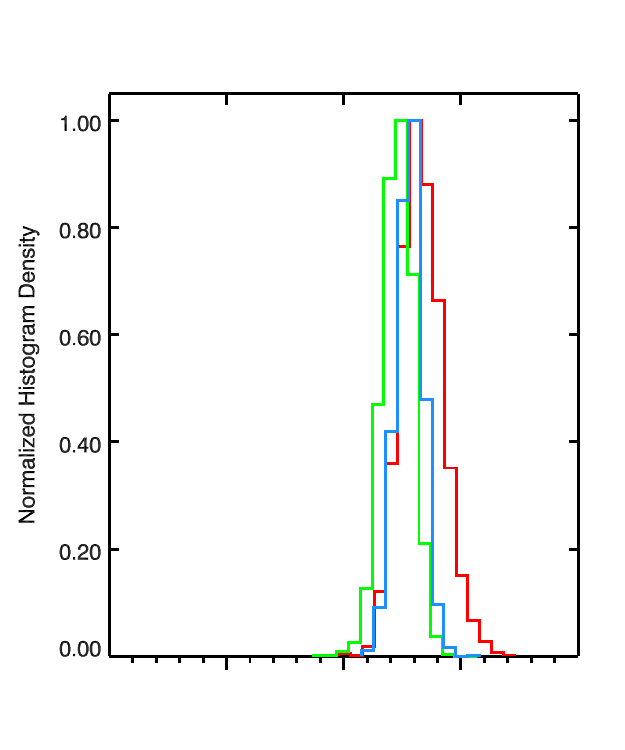}
\includegraphics[trim=0.4in 0.33in 0.0 0.35in, clip, height=0.18\textheight,width=0.3\textwidth]{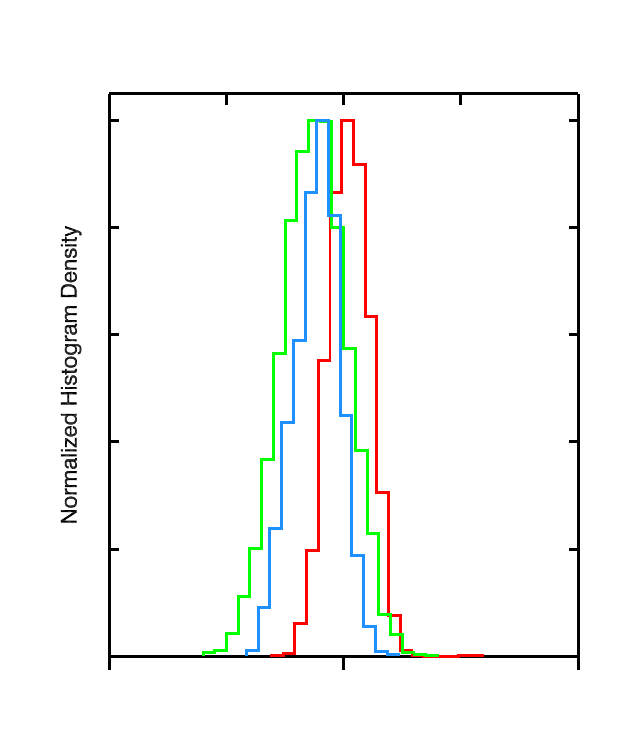}\\

\includegraphics[trim=0.0in 0.0in 0.0in 0.1in,clip,height=0.18\textheight,width=0.3\textwidth]{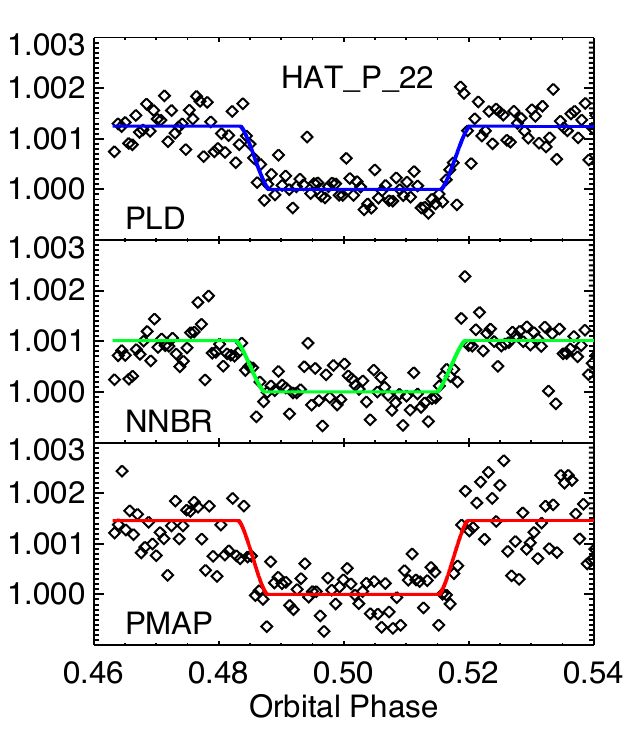} 
\includegraphics[trim=0.4in 0.0in 0.15in 0.35in,clip, height=0.18\textheight,width=0.3\textwidth]{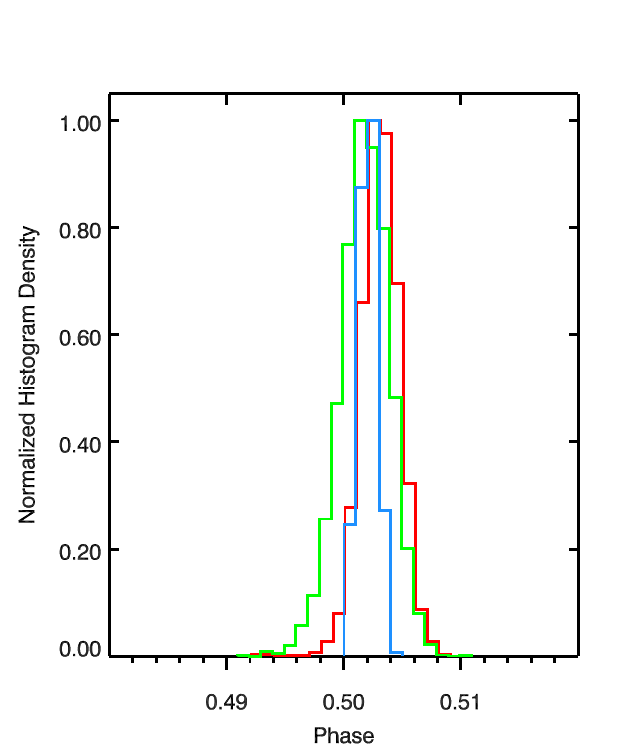}
\includegraphics[trim=0.4in 0.0in 0.0 0.35in, clip, height=0.18\textheight,width=0.3\textwidth]{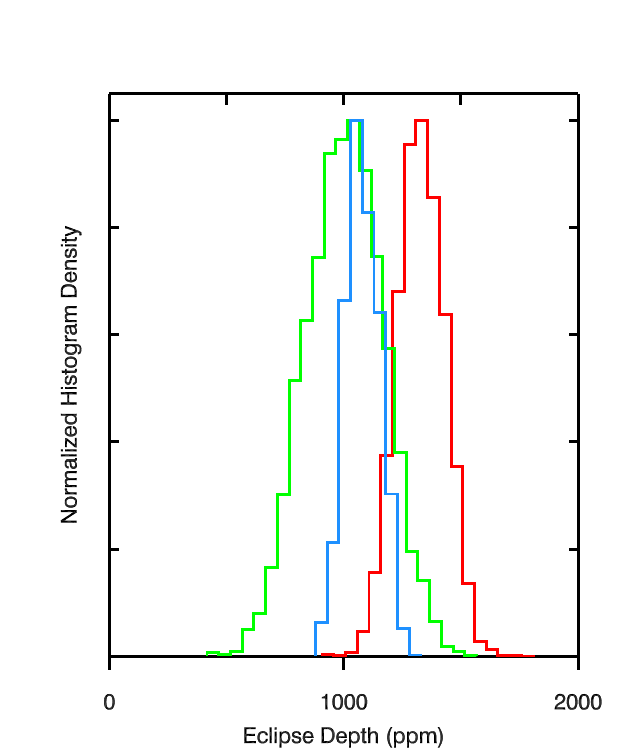}\\

\caption{The corrected photometric time series data and fit are shown in the left panels.  The right panels show the histograms of MCMC results for center of eclipse and eclipse depth.  PMAP (Red), NNBR (Green), and PLD (Blue).}\label{fits}
\end{figure*}

\section{Discussion}

\subsection{Systematic Implications}

\begin{figure}
\includegraphics[trim=0.65in 0.5in 0.2in 0.25in, clip, width=2.0 in, height=1.1 in]{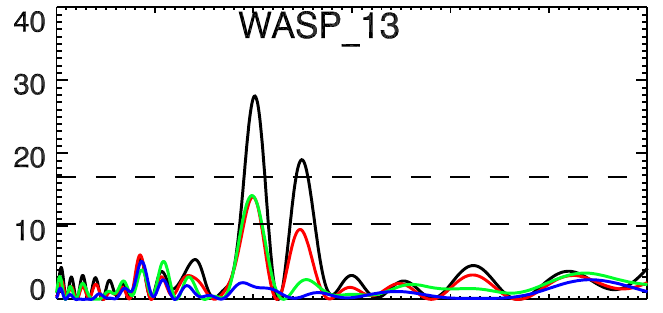}
\includegraphics[trim=3.0in 1.0in 2.0in 1.5in, clip, width=1.1 in, height=1.1 in]{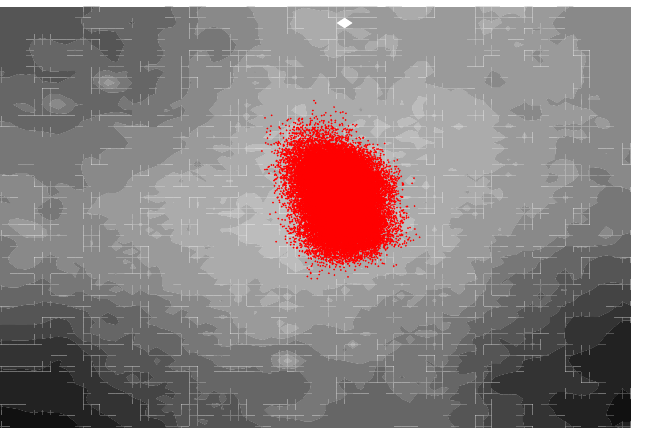}\\

\includegraphics[trim=0.65in 0.5in 0.2in 0.25in, clip, width=2.0 in, height=1.1 in]{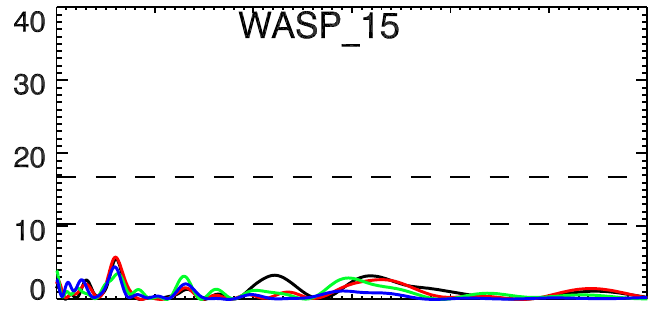}
\includegraphics[trim=3.0in 1.0in 2.0in 1.5in, clip, width=1.1 in, height=1.1 in]{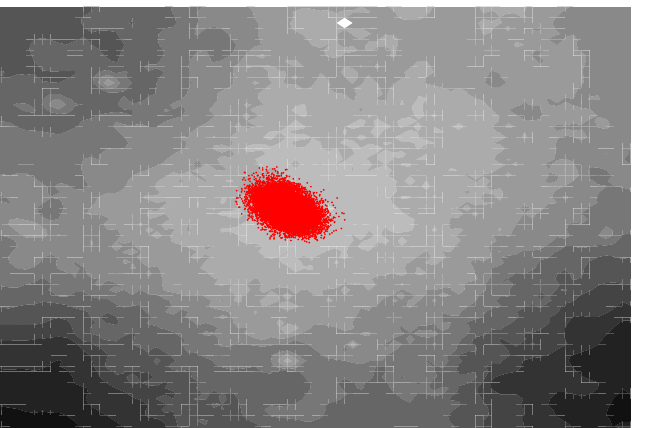}\\

\includegraphics[trim=0.65in 0.5in 0.2in 0.25in, clip, width=2.0 in, height=1.1 in]{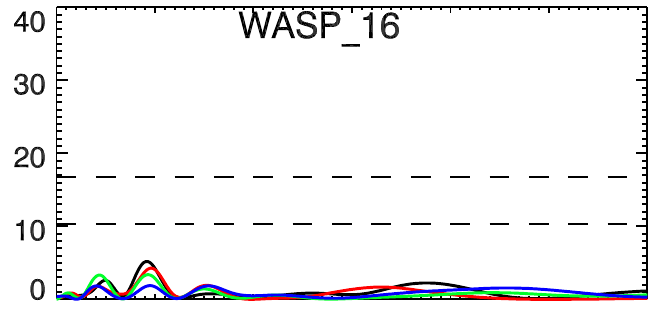}
\includegraphics[trim=3.0in 1.0in 2.0in 1.5in, clip, width=1.1 in, height=1.1in]{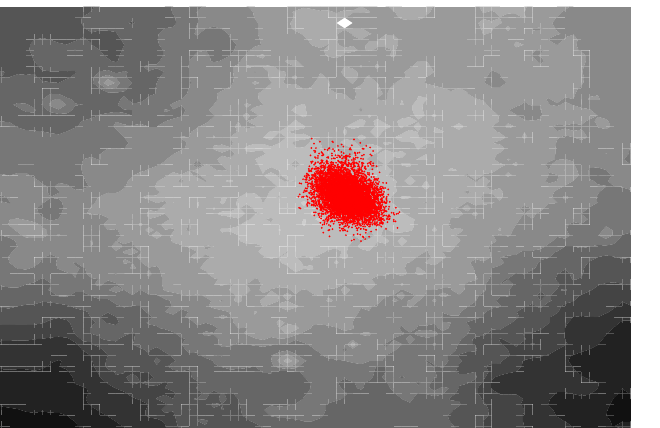}\\

\includegraphics[trim=0.65in 0.5in 0.2in 0.25in, clip, width=2.0 in, height=1.1 in]{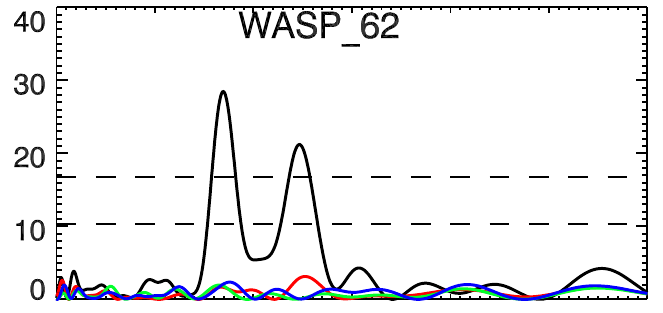}
\includegraphics[trim=3.0in 1.0in 1.5in 2.15in, clip, width=1.05 in, height=1.1 in]{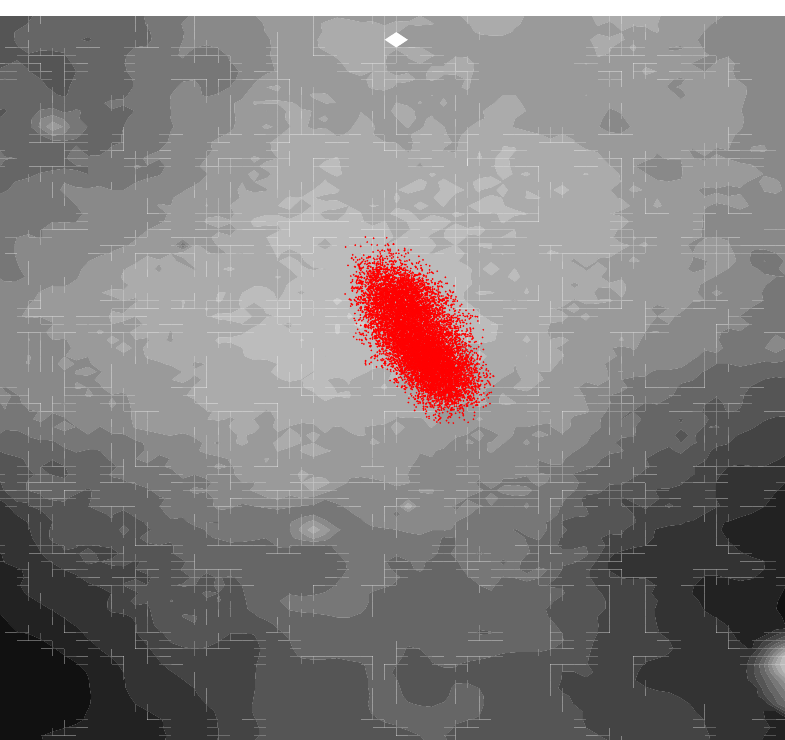}\\

\includegraphics[trim=0.65in 0.0in 0.2in 0.25in, clip, width=2.0 in, height=1.6 in, valign=t]{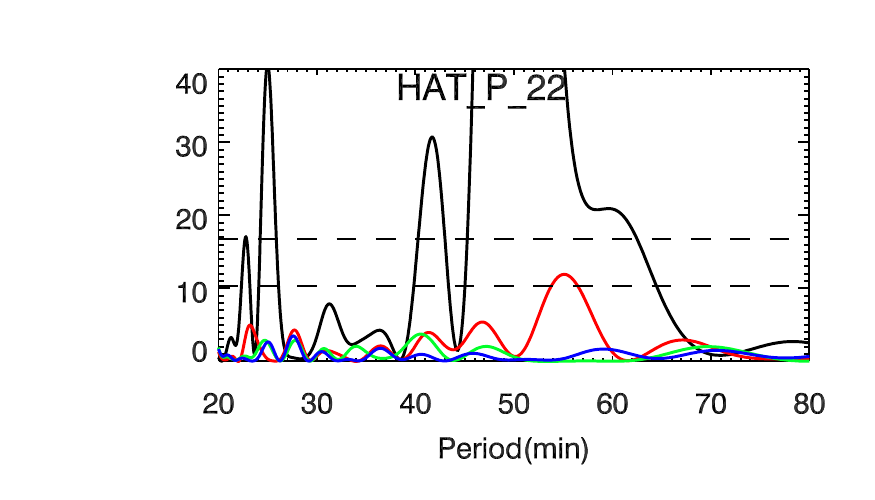}
\includegraphics[trim=3.0in 1.0in 2.0in 1.5in, clip, width=1.1 in, height=1.1in, valign=t]{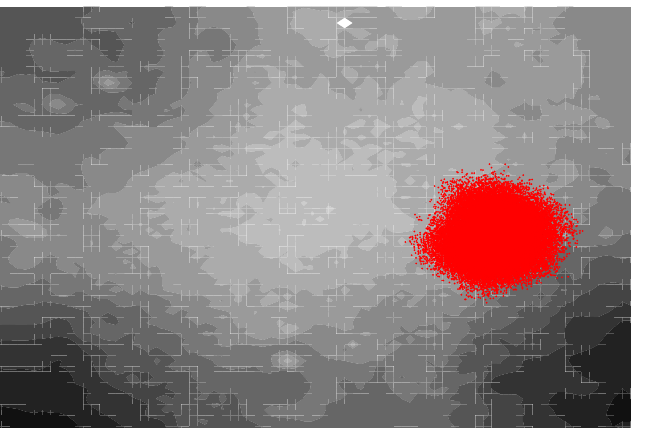}\\

\caption{LEFT:  Periodograms of the normalized power in the residuals of the uncorrected data (black), along with the periodogram of the residuals from fitting corrected data for each method (PMAP-red, NNBR-green, PLD-blue).  Significance levels 0.5 and 0.001 are shown as horizontal dashed lines.  RIGHT:  Positions of the stellar centroid overlaid on a gray scale image of the detector sweet spot as shown in Figure 1.  The period of the heater cycling responsible for most of the y drift is on the order of 40 minutes. Significant spikes in power are seen at this frequency in all but the most well positioned observations. }\label{scargle}
\end{figure}

As a primary goal, we sought to verify whether the application of each of these data analysis methods would produce statistically consistent results.  As shown in Table \ref{res} and the plots in Figure \ref{3plots}, this is generally true.  In all cases the NNBR and PLD methods agree to within $1\sigma$ uncertainty.  In all but two cases PMAP also agrees to within $1\sigma$ uncertainty.  The PMAP is reliant on pointing stability to keep the image on the well characterized sweet spot.  Examination of the centroid positions of these discrepant observations suggest that the image was either not on the most well characterized part of the detector and/or had significant $y$ drift in comparison to other observations.  The plots in Figure \ref{scargle} show the stellar centroid positions for each observation.

\capstartfalse
\def\arraystretch{1.5}
\tabletypesize{\small}
\begin{deluxetable}{lcccc}
\tablecaption{Secondary Eclipse Depth Results Using KMAP\label{nmapres}}
\tablewidth{0.45\textwidth}
\tablehead{
\colhead{Planet} & \colhead{Eclipse Depth} & \colhead{SDNR} &  \colhead{$\beta_{\rm red}$} & \colhead{Aperture Radius} \\
& \colhead{(ppm)} & & & \colhead{(pixels)}
} 
\startdata

WASP-13b             & 798 $\pm$ 193                                     & 0.0127                                          & 1.64                                 & 2.25                           \\ 
WASP-15b              & 1110 $\pm$ 216                                    & 0.0066                                          & 2.36                                 & 2.50                            \\
WASP-16b               & 1004 $\pm$ 128                                    & 0.0062                                          & 1.06                                 & 2.00                             \\ 
WASP-62b               & 1279 $\pm$ 110                                    & 0.0046                                          & 1.58                                 & 2.50                           \\ 
HAT-P-22b              &      999 $\pm$ 99                                               &     0.0069                                            &            1.62                          &           2.50                      
 \enddata
\end{deluxetable}
\capstarttrue

It is also worth noting that both of the data sets where PMAP did not match the NNBR and PLD reductions at the one sigma level utilized 0.4 s exposure times in contrast with the others that used 2.0 s exposure times.  It is entirely possible that this is purely coincidental, however, further investigation into differences in PMAP performance between targets of different brightness may be worthwhile.  Inspection of the centroid position plots in Figure \ref{scargle} suggest that the degrading performance efficiency with PMAP is more likely due to the drift in the stellar centroid.

The same team that developed PMAP has been recently employed a modification to their correction method that would utilize a kernel regression to create a calibration map (similar to NNBR) rather than the pre-gridded gain map \citep{2016Krick}.  We applied this new Kernel Regression with Pixel Map (KMAP),  to all of our targets as another means of comparison.  Each data set was corrected using KMAP with an array of fixed aperture sizes [2.00, 2.25, 2.50, 2.75, 3.00].  Similar to previous analyses, we chose the photometry that produced the fit which minimized the white and red noise.  In cases where stellar centroid is stable on the sweet spot, we find that results are not significantly different from the original PMAP results. However, the uncertainty in each result is reduced through improvements in both white and red noise. When we applied KMAP to the two cases where PMAP results were not within $1\sigma$  of NNBR and PLD we find that KMAP results are in close statistical agreement with PLD and NNBR ($<1\sigma$).  Results are shown in Table~\ref{nmapres}.

We also aimed to determine which methodologies produced the least amount of uncertainty in the eclipse depth and center of eclipse times.  In order to facilitate this evaluation we determine the white ($\sigma_w$) and red ($\sigma_r$)  components of the noise as described in \cite{Gillon2008} for comparison.  The white noise component is simply a measure of the scatter in the residuals after applying the best fit to the corrected data.  We can benchmark the white noise against the photon noise limit, and we can expect to achieve precision of approximately 1.3 times the photon noise limit using PMAP\footnote{\url{http://irsa.ipac.caltech.edu/data/SPITZER/docs/irac/warmimgcharacteristics/}}.  We find that the SDNR is, on average, 1.28, 1.15, and 1.15 times the photon noise limit for PMAP, NNBR, and PLD respectively.  
A cursory examination of the uncertainty associated with each eclipse depth in Figure \ref{fits} may lead one to believe that PMAP far out performed NNBR in limiting the scatter in the residuals.  However, this evidence would be counter to the $\sigma_w$ values which suggest that NNBR does significantly better than PMAP at reducing uncertainty.   Despite this, results from the MCMC still suggested that the NNBR results had a larger uncertainty.  What we find is that in the PMAP method the gain map itself is a fixed value map assumed to be correct (i.e. have no uncertainty) while the NNBR `map' is an additional degree of freedom, recalculated at every iteration of the MCMC. The pixel gain map does have an uncertainty $\sigma_P (x,y)$ as defined in \cite{Ingalls}, however, the correction routine provided by IPAC does not use the uncertainty in the map to scale the error in the corrected data.  The PMAP uncertainties given by the MCMC are underestimated as a result of this failure to propagate uncertainty forward from the gain map itself.  To verify, we fixed the NNBR map values after the LM fitting and passed it to the MCMC as a fixed parameter. This produced uncertainties substantially smaller than when it is allowed to vary and substantially smaller than the PMAP uncertainty (shown in Table \ref{res} denoted with the subscript `fixed').  We performed a similar test of PLD by fixing the coefficients of the pixel values ($c_i$) when passing to the MCMC rather than allowing the MCMC to treat them as free parameters.  This did not have a significant effect on the PLD results.  The uncertainties associated with NNBR and PLD are thought to be accurate and conservative estimates while the uncertainties of PMAP measurements are, perhaps, underestimated.

Another important measure of how an analysis method succeeds in reducing noise in \Spitzer time series data is how it reduces the correlated (red) noise.  As previously discussed in \autoref{sec: Intro} the source of the intrapixel sensitivity variation, the primary source of correlated noise in IRAC data, is the coupling between intrapixel gain variations and spacecraft pointing fluctuations.  A full correction for the effect would produce corrected data absent of any spatially correlated periodicity.  Figure \ref{scargle} shows the periodograms of the normalized power in the residuals of both the corrected and uncorrected data.  The heater cycle with a period of $\sim$40 minutes is the main source of spatially correlated noise.  We see power removed from this frequency when applying corrections to the data.  Any remaining power is on the order of random white noise and is not reduced if it is not spatially correlated. We measure the correlated noise remaining in the data after correction using both the $\beta_{\rm red}$ coefficient and $\sigma_{\rm red}$  \citep{Gillon2010}.  We find that the mean red noise component is 7.5\%, 6.0\%,  and 4.9\% of the magnitude of the white noise for PMAP, NNBR, and PLD respectively. 

Computational efficiency is another factor worthy of consideration in comparing reduction methods.  We will need to employ efficient methods to keep pace as the amount of data available increases. There are still a large number of observations to be analyzed, which were considered problematic before the development of the reduction techniques discussed in this work, and we anticipate another round of \Spitzer observations in the coming year.   Applying these methodologies to phase curves will require careful management of computational resources.  Phase curve observations are several times longer than eclipse observations making cumbersome methodologies impractical.

PMAP is by far the quickest way to reduce \Spitzer data and perform a fit.  The photometry of a single eclipse observation using a single aperture can be extracted and fit in a matter of minutes with this method.  

NNBR an PLD are very similar in their computational time.  Both make use of multiple fixed and variable apertures, effectively performing the reduction several times over.  PLD takes advantage of binning in order to speed up many of its calculations.  Fitting and performing $\chi^2$ analysis on binned data is far faster than working with the full data set.  The use of DeLauney Triangles to sort the nearest neighbors in the NNBR method greatly improves efficiency \citep{Lewis}. The number of neighbors utilized has been determined based upon the cost of computational time weighed against the improvement in SDNR.

The KMAP is the most time consuming method at this stage.  Both NNBR and KMAP recalculate a calibration map for each set of aperture photometry.  This is the most time consuming task in the process.  Further development of KMAP to find more efficient ways to perform the sorting and identification of the neighbors could lead to significant increases in computational efficiency.    

The advantage to using either the PMAP or KMAP methods is that they are not self calibrated.  Both NNBR and PLD rely solely on the data set at hand in calculating the corrections.  This makes them susceptible to degeneracies between the variation due to intrapixel sensitivity and the astrophysical signal.  Both PMAP and KMAP reference a grid of data points from a number of different calibration observations.  PMAP simply interpolates to find a solution based on the closest grid points while KMAP uses a kernel regression of nearest neighbors.  

\subsection{Atmospheric Implications}

\begin{figure}[!h]\label{phase}
\centering
\includegraphics[width=0.48\textwidth, height=0.35\textheight]{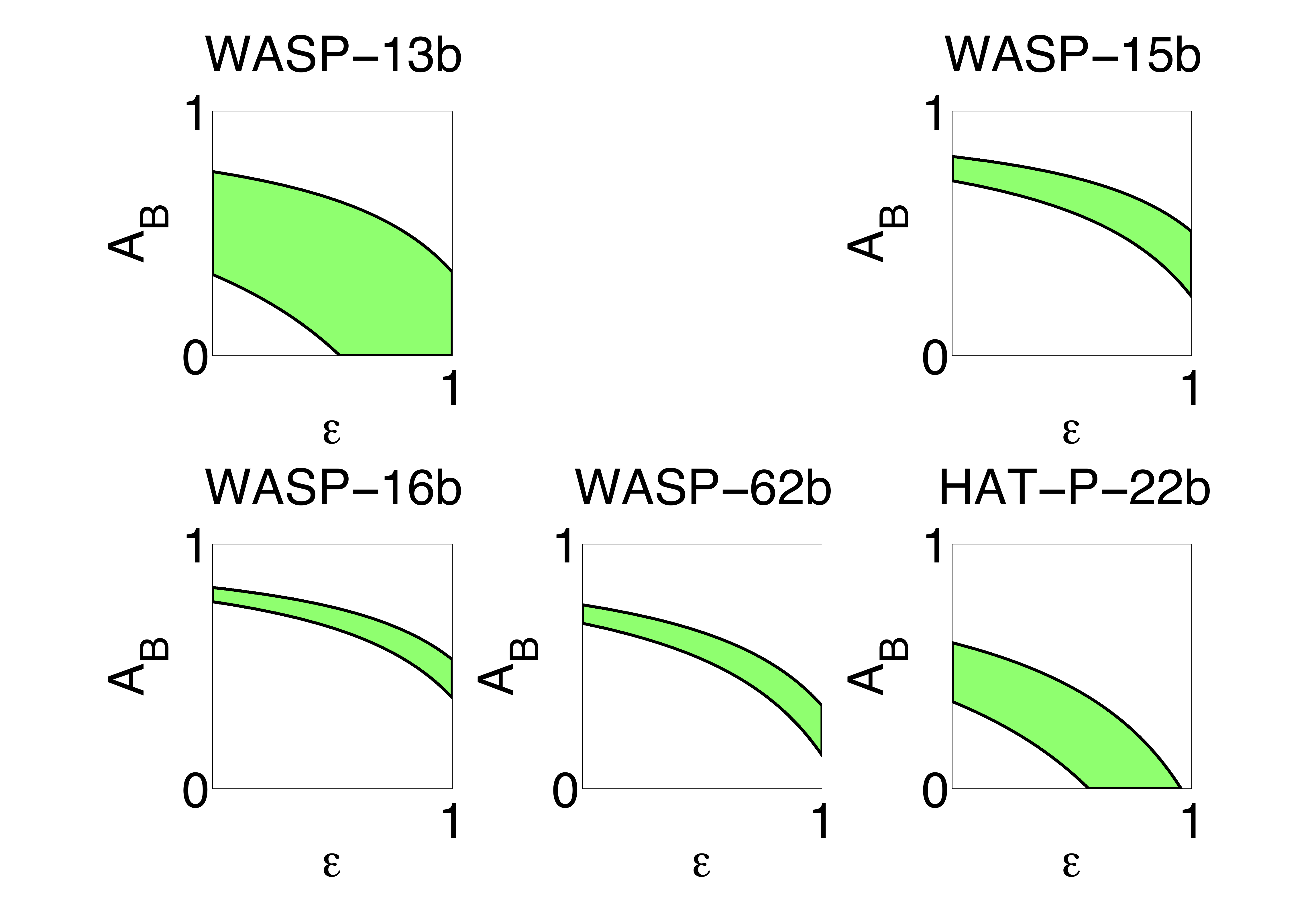}

\caption{Possible albedo ($A_B$) and recirculation efficiency ($\epsilon$) values that produce a dayside brightness temperature within the uncertainties of the weighted average brightness temperatures derived from our Spitzer $4.5\,\mu$m eclipses.  Calculations were performed using the methodology of \cite{2015SC}. Both $A_B$ and $\epsilon$ range from 0 to 1.  0 $A_B$ is total reflection and 1 full absorption. $\epsilon$ of 0 is no recirculation and $\epsilon$ of 1 is uniform redistribution.}
\end{figure}

These observations add five important data points to the exoplanets with measured infrared eclipses.  To constrain the atmospheric properties of our planet sample within the larger hot Jupiter population, we have calculated the $4.5\,\mu$m brightness temperatures, $T_B$, corresponding to each method in Table \ref{res}.  We then calculated a weighted-average $T_B$ to constrain each planet's bond albedo and recirculation.  We follow the methodology of \cite{2015SC}, who estimate a range of dayside effective temperatures ($T_{d}$) as
\begin{equation}\label{tday}
T_d=T_0(1-A_B)^{\frac{1}{4}}\left(\frac{2}{3}-\frac{5}{12}\epsilon\right)^{\frac{1}{4}}.
\end{equation}
Here T$_0$ is the equilibrium temperature. T$_0$ $\equiv T_{\rm eff *}\sqrt{\tfrac{R_*}{a}}$, where $T_{\rm eff *}$ is the stellar effective temperature, $R_*$ is the stellar radius, and $a$ is the orbital distance. Additionally, $A_B$ is the albedo, and $\epsilon$ is the recirculation efficiency, where $\epsilon=0$ implies no heat redistribution (i.e, no heat is transported from the dayside to the nightside), and $\epsilon=1$ implies full recirculation.  Because each brightness temperature is derived from a single eclipse, we can only constrain each planet's dayside temperature.  Therefore, our solution is degenerate, and a range of albedos and recirculation efficiencies are consistent with our measurements.  We perform a $\chi^2$ analysis using a weighted-average of the brightness temperatures over a $101\times 101$ grid of albedo and recirculation values as was done in \cite{2015SC}, and interpolate the median values.  Our results are shown in Fig. \ref{phase}.  Furthermore, we compare our average $T_B$ to upper and lower bounds of equilibrium temperatures in Table \ref{temps}.  The lower bound is calculated assuming an albedo of 0.4 and full recirculation ($\epsilon=1$) while the upper bound is calculated assuming an albedo of 0.0 and no recirculation ($\epsilon=0$). 

\capstartfalse
\def\arraystretch{1.5}
\tabletypesize{\footnotesize}
\begin{deluxetable}{lccc}
\tablecaption{Comparison of the weighted-average brightness temperature (T$_B$) with upper and lower bounds represented by T$_{\rm eq 1}$ and T$_{\rm eq 2}$ all in $^\circ $K.   T$_{\rm eq 1}$ is calculated as a lower bound using an albedo of 0.4 and uniform recirculation ($\epsilon=1$), and T$_{\rm eq 2}$ is an upper bound with albedo of 0 and $\epsilon$ of 0.  \label{temps}}
\tablewidth{0.45\textwidth}
\tablehead{
\colhead{Planet} & \colhead{T$_{\rm eq 1}$} & \colhead{T$_{\rm eq 2}$} &  \colhead{Avg. T$_B$}  } 
\startdata

WASP-13b &  1365.4     &   1983.1      & 1595.1$\pm$197.9       \\ 

WASP-15b        & 1452.1        & 2108.41       & 1461.1$\pm$79.1     \\ 

WASP-16b        & 1149.7         & 1669.3        & 1122.7 $\pm$40.5     \\ 

WASP-62b        & 1254.6        & 1821.6        & 1329.6$\pm$44.8        \\ 

HAT-P-22b       & 1126.5         & 1635.6        & 1383.9$\pm$81.3      \\ 

\end{deluxetable}
\capstarttrue

Comparing each map in Figure \ref{phase}, we see that WASP-13b has the largest possible range of albedos and recirculation efficiencies; this is a result of its large uncertainty in weighted-average brightness temperature (Table \ref{temps}).  If we compare the weighted-average brightness temperatures to the upper and lower bound equilibrium temperatures, we see that the brightness temperatures for WASP-13b, WASP-15b, WASP-16b and WASP-62b are nearer in value to the lower bound equilibrium temperature, suggesting that those planets' atmospheres favor moderate albedo and high recirculation.  All of these targets have orbital periods on the order of several days, and the $\frac{a}{R_*}$ values are greater than 7. Previous works have shown that planets in this regime are far more likely to have efficient recirculation \citep{pbs2013,cowanagol2011,2015SC,2015Kammer}.

The weighted-average brightness temperature of HAT-P-22b is more centered between the two extremes, suggesting the atmosphere may favor moderate albedo and less efficient recirculation.  Because the atmosphere is expected to be dominated by equatorial super-rotation at photospheric pressures \citep{showbook}, this might suggest other atmospheric processes at play on HAT-P-22b to inhibit recirculation.  For example, it is possible that the infrared photosphere could be located at lower pressures than we expect, where radiative timescales are short compared to advective timescales. This could result from a low C/O ratio, which would imply an excess of CO and produce enhanced opacity at 4.5 μm where CO has strong vibrational bands \citep{2015Wong}. Enhanced opacity could also arise if the atmospheric metallicity were greater than solar (e.g. Lewis et al. 2010; Kataria et al. 2015); this could also produce a dayside temperature inversion, which would additionally move the IR photosphere to lower pressures. Clearly, multi-wavelength eclipse and phase-curve measurements are needed for all five planets to further constrain the radiative, advective, and chemical processes taking place in their atmospheres.

\section{Conclusions}

We present secondary eclipse measurements for WASP-13b, WASP-15b, WASP-16b, WASP-62b,  and HAT-P-22b in IRAC channel 2 at 4.5 $\mu$m.
The reduction of \Spitzer photometric observations requires accurate corrections to the intrapixel sensitivity effect in order to achieve the precision necessary to extract eclipse depths.  The methodologies discussed in this work all are suitable approaches for achieving this goal. The computational efficiency and reliability with which PMAP is able to produce an eclipse fit makes it a valuable tool which researchers can use to quickly produce results.  Each of the other methods considered potentially improve upon PMAP in producing a more precise measurement by their handling of both correlated and uncorrelated noise effects. However, this improvement comes at the cost of computational efficiency.   PMAP proves quite reliable in cases where the peak up method of observation has been employed and the stellar centroid is well positioned on the detector.  We have shown that methods that employ a nearest neighbor kernel handle cases where centroid position or drift may not be optimal. PLD may not be as well suited for data sets with large a amount of drift \citep{2015Wong}, however, the observations considered here are relatively short and stable in comparison to phase curves and PLD performed quite well.  Both the PLD and NNBR methods consistently minimized the correlated noise in the corrected data.  

We find that the brightness temperatures derived for WASP-13, WASP-15, WASP-16, and WASP-62 from our measured eclipse depths place them in a regime requiring either high albedo or efficient recirculation.  While it is possible that these these planets have a higher albedo than previously observed planets, we suggest that there are a myriad of other physical processes not considered in a simple albedo-recirculation model that would produce similar results.  HAT-P-22b occupies an area of phase where moderate albedo (0.4) and much less efficient recirculation would produce the measured brightness temperature.  
\section{Acknowledgments}

BMK acknowledges Brown University for its financial support of his contributions to this work.  This work is based on observations made with the {\sl Spitzer Space Telescope}, which is operated by the Jet Propulsion Laboratory, California Institute of Technology under a contract with NASA.  We also acknowledge that part of this work was completed at the Space Telescope Science Institute (STScI) operated by AURA, Inc.  This research has made use of the Exoplanet Orbit Database
and the Exoplanet Data Explorer at exoplanets.org.  

\bibliography{spitzer3}


\end{document}